\documentclass[lettersize,journal]{IEEEtran}
\usepackage{amsmath,amsfonts}
\usepackage{algorithm}
\usepackage{array}
\usepackage[caption=false,font=normalsize,labelfont=sf,textfont=sf]{subfig}
\usepackage{textcomp}
\usepackage{stfloats}
\usepackage{url}
\usepackage{verbatim}
\usepackage{cite}
\usepackage{booktabs}
\usepackage{graphicx}
\usepackage{algpseudocode}
\usepackage{caption}
\usepackage{titlesec}
\usepackage{comment}
\newcommand{\ScatteringTransform}[1]{\text{Scattering Transform of } #1}

\excludecomment{exclude}

\begin{document}

\title{Integrating Graph Neural Networks with
Scattering Transform for Anomaly Detection}

\author{%
\IEEEauthorblockN{Abdeljalil Zoubir\textsuperscript{a,*}, Badr Missaoui\textsuperscript{b}} \\
\IEEEauthorblockA{\textsuperscript{a}College of Computing, UM6P, Lot 660, Ben Guerir, 43150, Morocco} \\
\IEEEauthorblockA{\textsuperscript{b}Moroccan Center for Game Theory, Rabat,10000, Morocco} \\
\thanks{* Corresponding author }
\thanks{Email adresses:
abdeljalil.zoubir@um6p.ma (Abdeljalil ZOUBIR) 
badr.missaoui@um6p.ma (Badr MISSAOUI)}
}

\maketitle

\begin{abstract}
In this paper, we present two novel methods in Network Intrusion Detection Systems (NIDS) using Graph Neural Networks (GNNs).
The first approach, Scattering Transform with E-GraphSAGE (STEG), utilizes the scattering transform to conduct multi-resolution
analysis of edge feature vectors. This provides a detailed representation that is essential for identifying subtle anomalies in network
traffic. The second approach improves node representation by initiating with Node2Vec, diverging from standard methods of using
uniform values, thereby capturing a more accurate and holistic network picture. Our methods have shown significant improvements
in performance compared to existing state-of-the-art methods in benchmark NIDS datasets. 
\end{abstract}


\section{Introduction}

\IEEEPARstart{C}{ombating} the increasing frequency and complexity of cyber attacks on computer networks requires a comprehensive strategy integrating technological, procedural, and educational aspects. Organizations adopt diverse methods to address these challenges, such as installing firewalls, implementing multi-factor authentication, conducting regular security assessments, and educating staff on cybersecurity principles. Network Intrusion Detection Systems (NIDS) and Network Intrusion Prevention Systems (NIPS) play a crucial role in this defense, offering real-time surveillance and active defense capabilities.
While NIDS have been fundamental in cybersecurity, they present challenges, including generating false positives and negatives. Encrypted communications further complicate their effectiveness, concealing malicious activities. The surge in network traffic strains these systems, and their reliance on signature-based detection methods exposes them to new cyber threats.

 To overcome these challenges, there is a growing focus on integrating graph-based approaches that consider topological patterns. By exploiting the inherent structure of network traffic,

graph-based methods can offer a more comprehensive view, capturing relationships and interactions that traditional NIDS may miss. This approach not only reduces false positives and negatives but also excels in identifying complex attack patterns within network traffic, thus providing a more robust defense against evolving cyber threats.
 
 For anomaly detection, where relational context is paramount, Graph Neural Networks (GNNs) provide superior representations by seamlessly integrating node and edge attributes with the inherent graph structure. This holistic approach enhances their ability to detect nuanced patterns, positioning GNNs as particularly adept at identifying anomalies that conventional models may miss in large, interconnected datasets.

In our study, we present a novel approach to NIDS by exploiting the capabilities of GNNs combined with advanced feature extraction techniques. Recognizing the critical role of edges in network analysis,  Our first method employs the scattering transform, enabling us to analyze edge feature vectors at different resolutions. It helps us to capture both granular interactions and broad trends. On the other hand, nodes in graph-based studies often lack inherent properties, resulting in an unbalanced representation. Common strategies typically initialize these nodes with uniform values and updating them later based on adjacent edge features, but this may fail to capture the latent dynamics of the network. To overcome this challenge, our strategy incorporates Node2Vec \cite{n2v}. This technique maintains both localized and overarching network structures, enriching the nodes’ initial representation. Consequently, Node2Vec sets the stage for a more comprehensive and informed analysis of the graph.

In graph-based learning, as highlighted in references \cite{hamilton2017inductive}, a significant focus has been on node or topological features, often employing random walk strategies. This approach, though effective in certain contexts, can neglect essential edge information, which is indispensable for NIDS. Moreover, focusing solely on edge data without delving into the intrinsic, multi-layered information within them, as noted in \cite{caville2022anomal}, does not fully exploit their analytical potential. Our approach, STEG, marries the scattering transform with E-GraphSAGE \cite{lo2021graphsage} to thoroughly access and leverage this critical edge data. E-GraphSAGE, an extension of the original GraphSAGE \cite{hamilton2017inductive} model, is specifically designed to capture edge features and topological patterns in graphs. Concurrently, by integrating Node2Vec as initialization, our methodology ensures a robust representation of nodes, creating a comprehensive and balanced framework for both node and edge analysis within the E-GraphSAGE framework.

Our contributions are specifically designed for network intrusion detection without the need for data labels, utilizing self-supervised learning in two distinct approaches. The first approach, STEG, employs the Scattering Transform to extract deeper and more informative edge embeddings. This enables a nuanced understanding of network behavior.  In our second approach, we employ self-supervised learning through Node2Vec to extract valuable graph topology features. These methods collectively advance the field of Graph Neural Networks (GNNs) for network intrusion detection, offering a practical and meticulously analyzed solution.
 
As far as we are aware, our approaches outperforms existing self-supervised learning methods for anomaly detection in network intrusion. In summary, the key contributions of our paper are:

\begin{enumerate}
\item 
 Utilizing STEG, a self-supervised GNN-based approach, enhances network intrusion detection by leveraging the Scattering Transform to extract comprehensive information from edge features, without relying on labeled data. This provides a powerful and versatile solution for anomaly detection in network security. STEG employs a modified version of the E-GraphSAGE framework \cite{Lo2021EGraphSAGEAG}, integrating state-of-the-art self-supervised learning techniques for GNNs and utilizing traditional anomaly detection methods tailored for network intrusion detection. This marks the first use of self-supervised GNNs and the scattering transform in network security.

\item 
Unlike Anomal-E \cite{caville2022anomal} or the traditional E-GraphSAGE approach, which typically initializes node features with a uniform value, our alternative method adopts Node2Vec initialization. This approach initializes node features with real-world attribute values. imbuing each node with meaningful information from the start. This strategy aims to enhance model proficiency in identifying anomalies in network behavior, leading to greater precision and contextual awareness.

\item 
Two benchmark NIDS datasets for network intrusion detection serve as the testing grounds for the application of our dual approaches. Through extensive experimental evaluations, these approaches have shown significant enhancements in performance compared to existing state-of-the-art methods, highlighting their potential effectiveness in the field.
\end{enumerate}

The structure of this paper is outlined as follows: Section ~\ref{related} reviews related work, while Section ~\ref{Background} provides essential background on Graph Neural Networks and related concepts. Section ~\ref{anoml-E}. introduces our methodologies: Scattering Transform with E-GraphSAGE (STEG) and Node2Vec initialization with E-GraphSAGE, along with details on data preparation and our training approach. Section ~\ref{results} focuses on the implementation and refinement of Anomaly Detection Algorithms, followed by the presentation of experimental evaluation results. and Section \ref{conclusion} concludes the paper.

 \section{Related Work} \label{related}
 
As our work not only utilises cutting-edge methods in GNNs, but also integrates additional feature extraction techniques with E-GraphSAGE \cite{Lo2021EGraphSAGEAG}, we are pioneering the application of these advanced methods to network anomaly detection in a self-supervised framework. Given the innovative nature of our approach, there is a limited amount of closely related prior work. Therefore, the research discussed below is related to our approach in a broader sense, encompassing the fields of GNNs, feature extraction and self-supervised learning in network anomaly detection.

Zhou et al. \cite{Zhou} introduced an innovative GNN model for botnet detection in NIDS. Their approach prioritized network topology over individual node and edge characteristics. To simulate real-world conditions, authentic background traffic was incorporated into established botnet datasets. The model used topological data aggregation, using constant vectors for node characteristics and excluding detailed edge characteristics. This simplified the analysis of network structure and bypassed the complexity of node and edge attributes. Zhou et al. demonstrated superior performance compared to traditional methods such as logistic regression and BotGrep \cite{grep}. This highlights the potential of focusing on network topologies to improve the efficiency of botnet detection.

Caville et al. \cite{caville2022anomal} introduce a self-supervised GNN approach to NIDS, focusing on edge features to detect network intrusions and anomalies. This method shows significant progress, especially in adapting to sophisticated attacks. In their study, Anomal-E outperformed traditional methods such as GraphSAGE, with Anomal-E-CBLOF and Anomal-E-HBOS achieving macro F1 scores of 93.04\% and 91.89\%, respectively, significantly higher than GraphSAGE-CBLOF and GraphSAGE-HBOS at 72.07\% and 74.69\%. However, aspects of the Anomal-E methodology, such as initializing node features with constant vectors like \cite{Lo2021EGraphSAGEAG} and incorporating Deep Graph Infomax (DGI), may limit its effectiveness. The constant vectors could oversimplify the feature representation, which could affect anomaly detection in complex scenarios. In addition, DGI's focus on maximising information in the latent space of the graph may not match the unpredictable nature of network intrusions, potentially limiting the model's ability to detect a wide range of anomalies.

Layeghy et al.  \cite{siamak} conducted a comprehensive investigation into the performance and generalisation of ML-based NIDS, evaluating seven different supervised and unsupervised machine learning algorithms on four recently published NIDS benchmark datasets. The study included unsupervised learning experiments with Isolation Forest (IF), One-Class Support Vector Machines (oSVM), and Stochastic Gradient Descent One-Class Support Vector Machines (SGD-oSVM). The key finding of their research was that, in the context of these NIDS datasets, unsupervised anomaly detection algorithms generally demonstrated better generalisation capabilities than their supervised counterparts. This finding is particularly significant as it suggests that unsupervised learning methods may be more effective for network intrusion detection in certain scenarios, potentially providing a more robust and adaptable approach in diverse network environments.

Song et al. \cite{SONG2022108274} used self-supervised GNNs in the context of Knowledge Tracing (KT) and presented a distinctive approach with their Bi-Graph Contrastive Learning based Knowledge Tracing (Bi-CLKT) model. This innovative model focuses on capturing both local node-level details and global graph structures, using a unique training methodology based on contrasting graph pairs. By integrating node and graph embeddings, Bi-CLKT effectively navigates between detailed individual analysis and comprehensive graph-level understanding. The model's effectiveness is underlined by its significant outperformance against established KT models, demonstrating the broad applicability and effectiveness of self-supervised GNNs in complex, graph-based data environments.

Li et al. \cite{li2024controlled} introduce controlled graph neural networks (ConGNN) for network anomaly detection using a denoising diffusion probabilistic model (DDPM). ConGNN addresses data scarcity by generating augmented data through a graph-specific diffusion model, facilitating feature transfer between nodes to improve identification. Consisting of a DDPM-based embedding generator, a controlled GNN, and a hypersphere-based semi-supervised detection model, ConGNN demonstrates superior performance over baselines. However, its effectiveness can be hampered by its reliance on the selection of appropriate reference nodes for data augmentation, potentially impacting accuracy in complex network environments.

Alshammari et al. \cite{ALSHAMMARI2022100192} propose a parameter-free method for spectral clustering and SpectralNet to efficiently detect clusters with non-convex shapes. This innovative approach bypasses the need for parameter tuning by adaptively filtering neighbors based on surrounding density, retaining only highly similar connections. Validated on synthetic and real-world datasets, it provides a stable and computationally efficient alternative. However, it does not reduce the number of graph vertices, which may affect its efficiency on large datasets. Future improvements could include incorporating a vertex reduction component sensitive to local statistics or exploring alternative kernels for similarity computations beyond Gaussian.

The Deep Adversarial Anomaly Detection (DAAD) method by Zhang et al. \cite{zhang2022deep} marks a significant improvement in anomaly detection, combining self-supervised learning with adversarial training. While it effectively addresses key challenges in conventional methods, its complexity and high computational demands may restrict its use in limited-resource settings. Additionally, DAAD's performance heavily depends on data quality, and it requires extensive tuning, potentially limiting its applicability in diverse or rapidly evolving scenarios. Despite its efficacy in specific datasets, the generalization of DAAD across various domains is yet to be fully established.

\section{Background}\label{Background}
Graph-based representation learning is being more acknowledged as a transformational avenue of modern research, with applications ranging from medical diagnoses to molecular structures. GNNs have established themselves as cutting-edge tools in the cybersecurity scene, notably in network intrusion detection, as seen by their success in detecting sophisticated cyber threats such as Anomal- E \cite{caville2022anomal}. 

\subsection{Graph Neural Networks}
Graph Neural Networks (GNNs), a breakthrough in deep learning, are tailored for structured data analysis \cite{zhou2020graph}. Unlike traditional neural architectures such as Convolutional Neural Networks (CNNs), GNNs excel in processing complex data linkages inherent in graph structures.

GNNs leverage 'message passing' to aggregate information from neighboring nodes, enabling a holistic understanding of each node's context \cite{wu2020comprehensive}. This results in embeddings, condensed vector representations useful for tasks like node categorization and anomaly detection.

In cybersecurity, GNNs prove invaluable for analyzing intricate computer networks, where hosts represent nodes and interactions form edges. GNNs excel in navigating this data complexity, adept at identifying elusive security threats like Zero-Day Exploits \cite{bilge2012before}.

\subsection{GraphSAGE}\label{GraphSAGE Algorithm}

\textit{GraphSAGE}, developed by Hamilton et al., is a pivotal technique in GNNs \cite{Hamilton2017}. It introduces inductive learning, enabling node embeddings not only in static graphs but also in dynamic scenarios. Central to GraphSAGE is its neighbor sampling mechanism, efficiently propagating messages by selectively sampling node neighbors. This optimized sampling ensures computational efficiency across various graph structures and sizes.

GraphSAGE operates on a graph $\mathcal{G}(\mathcal{V}, \mathcal{E})$, where $\mathcal{V}$ represents nodes, $\mathcal{E}$ represents edges, and $x_v$ denotes node features. The $k$-hop neighborhood depth and aggregation function $AGG_k$, for $ k\in \{1,...,K\}$, is essential to GNN dynamics. GraphSAGE accommodates diverse aggregation strategies, including spectral-based convolution from Graph Convolutional Networks (GCN) \cite{Hamilton2017}, enhancing feature extraction and embedding robustness.

\subsubsection{Node Embedding process}
The process of collecting data from a nearby neighborhood of a node, similar to convolutional methods observed in CNNs, is integral to the computation of node embeddings. The GraphSAGE algorithm assumes that weight matrices and aggregator function parameters are predefined after training.

GraphSAGE iteratively collects data from a node's nearest neighbors. Each iteration begins by sampling the neighborhood of the node and merging the extracted data into a unified vector. Upon reaching the $k$-th layer, the aggregated information $\mathbf{h}_{N(v)}^{k}$ for a node $v$, rooted from its sampled neighborhood $N(v)$, is expressed by equation (\ref{eq:general_aggregator}) \cite{Hamilton2017}.
\begin{equation}\label{eq:general_aggregator}
    \mathbf{h}_{\mathcal{N}(v)}^{k} = \text { AGG }_{k}\left(\left\{\mathbf{h}_{u}^{k-1}, \forall {u} \in \mathcal{N}(v)\right\}\right).
\end{equation}

Breaking down the equation, $\mathbf{h}_{u}^{k-1}$ symbolizes the embedding of the node $u$ in the previous layer. The key to this process is aggregating the embeddings of all nodes, represented by $u$, that are near to $v$. This aggregated data then crafts the embedding for node $v$ when the algorithm is operating on the $k^{th}$ layer.

\begin{figure}[!t]
    \centering
        \includegraphics[width=0.95\columnwidth]{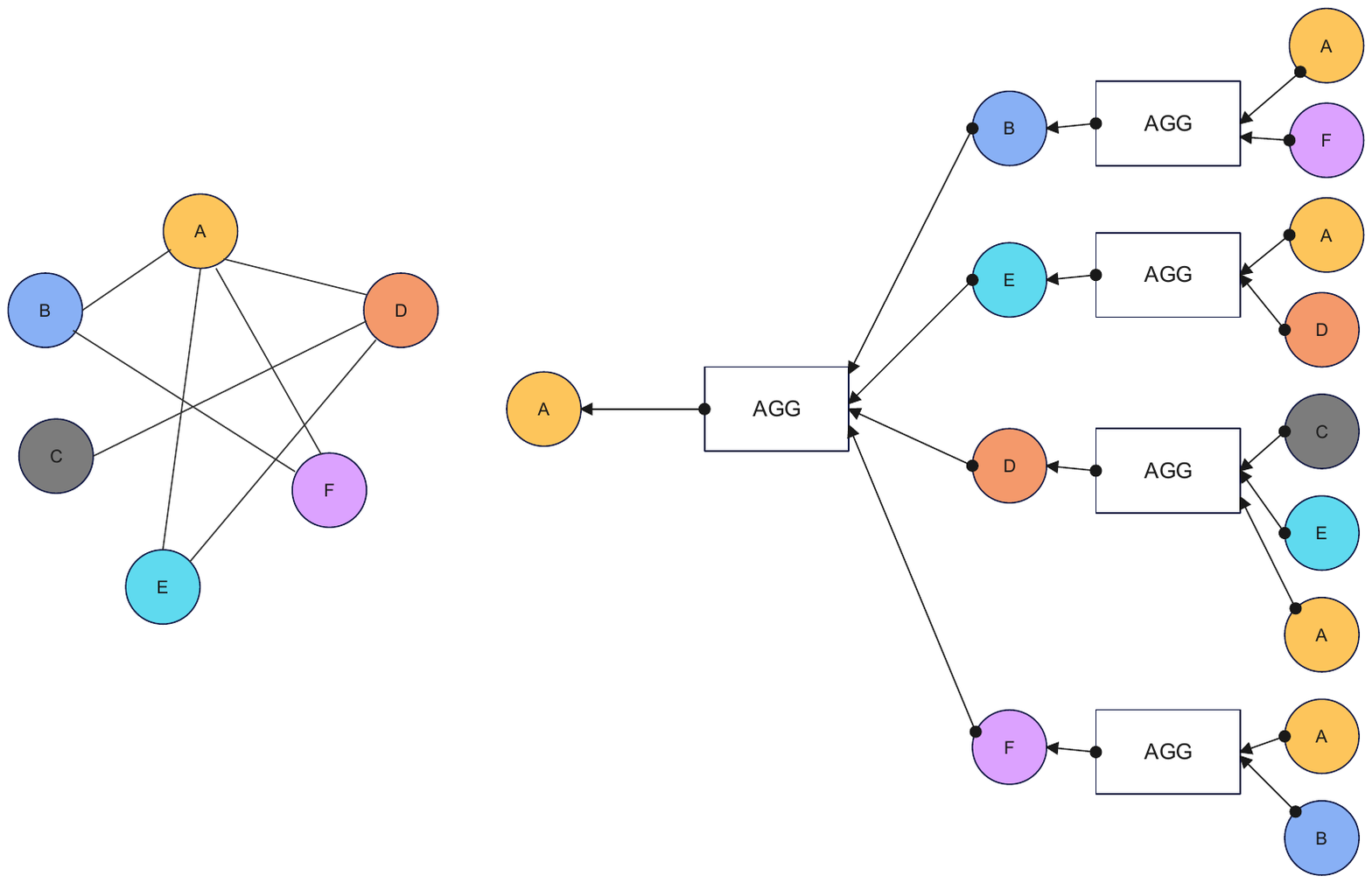}
    \caption{Illustration of a Basic Graph (left) juxtaposed with its GraphSAGE Representation using Two-Level Convolutions (right) encompassing Complete Neighborhood Sampling.} 
    \label{fig:graphsage_overview}
\end{figure}

The aggregation process, as shown in Figure \ref{fig:graphsage_overview} (right); takes into consideration the k-hop neighboring node properties of each node in the network, bringing them together using a specific aggregating procedure.
We build a comprehensive representation from the nodes sampled inside this scope, which is then fused with the node's prior layer representation, denoted as $\mathbf{h}_v^{k-1}$. Following that, the model applies its weight matrix $\mathbf{W}^k$ to this fusion. The result of this computation is then passed via a nonlinear activation function $\sigma$ (e.g., ReLU), resulting in the extraction of the node embeddings $\mathbf{h}_{v}^{k}$ for that specific layer, as defined in Equation (\ref{eq:general_graphsage}) \cite{Hamilton2017}:

\begin{equation}
 \small
 \label{eq:general_graphsage}
 \mathbf{h}_{v}^{k} = \sigma\left(\mathbf{W}^{k}\cdot\operatorname{CONCAT}\left(\mathbf{h}_{v}^{k-1}, \mathbf{h}_{\mathcal{N}(v)}^{k}\right)\right). 
\end{equation}

The ultimate representation for node $v$ is given by $\mathbf{z}_v$. This represents the node's embeddings derived at the terminal layer $K$, as elucidated in (\ref{eq:final_embedding}). In scenarios where node classification is essential, these embeddings (or node representations) denoted by $\mathbf{z}_v$, can be processed through either a sigmoidal neuron or a softmax layer to fulfill classification objectives.

 \begin{equation}
 \label{eq:final_embedding}
     \mathbf{z}_v = \mathbf{h}_v^K,\ \ \ \ \ \forall v \in \mathcal{V}.
 \end{equation}
 

\subsection{Enhanced E-GraphSAGE}\label{Edge-Based GraphSAGE}
The advancement of GNNs has revolutionized various fields. However, many approaches overlook edge characteristics in favor of node qualities for information dissemination. E-GraphSAGE addresses this gap by leveraging $K$-hop edge attributes to extract edge embeddings, particularly for network intrusion detection. This framework streamlines the creation of edge embeddings, aiding in distinguishing between benign and malicious network flows.

In network security, NIDS are crucial for identifying and blocking hostile network flows, akin to edge classification in graph-based representation learning. Historically, the GraphSAGE method prioritized node attributes over edge features \cite{Hamilton2017}, despite the prevalence of network flow data in attack datasets, leading to confusion.

To address this, the approach now emphasizes sampling and aggregating edge characteristics at a fixed k-hop depth, prioritizing edge embeddings for detecting malicious network activity. These enhancements led to the development of the E-GraphSAGE method, detailed in Algorithm~\ref{alg:E-GraphSAGE-Scattering}.

\begin{algorithm}[!t]
\caption{Generalized E-GraphSAGE edge and node embedding with Scattering Transform and enhanced initializations.}\label{alg:E-GraphSAGE-Scattering}
\textbf{Input:} \\
Graph $\mathcal{G}(\mathcal{V}, \mathcal{E})$; \\
edge features $\{\mathbf{e}_{uv}, \forall uv \in \mathcal{E}\}$; \\
initial node features $\mathbf{x}_v$ (could be constant vectors, Node2Vec, etc.), $\forall v \in \mathcal{V}$; \\
depth $K$; \\
weight matrices $\mathbf{W}^k$, $k \in\{1, \ldots, K\}$; \\
non-linearity $\sigma$; \\
aggregator functions ${AGG}_k$.

\textbf{Output:} \\
Edge embeddings $\mathbf{z}_{uv}, \forall uv \in \mathcal{E}$; \\
Node embeddings $\mathbf{z}_v$, $\forall v \in \mathcal{V}$.

\begin{algorithmic}[1]
\State Initialize: $\mathbf{h}_v^0 \leftarrow \mathbf{x}_v$, $\forall v \in \mathcal{V}$
\For{$k \leftarrow 1 \textbf{to} K$}
    \For{$uv \in \mathcal{E}$}
        \State Apply Scattering Transform on $e_{uv}$:
        \State $s_{uv}^k \leftarrow \ScatteringTransform{e_{uv}}$
        \State Update edge feature:
        \State $e_{uv}^k \leftarrow \operatorname{CONCAT}({e_{uv}^{k-1}, s_{uv}^k})$
    \EndFor
    \For{$v \in \mathcal{V}$}
        \State Aggregate information:
        \State $h_{\mathcal{N}(v)}^k \leftarrow AGG_k(\{h_u^{k-1}, e_{uv}^k\}, \forall u \in \mathcal{N}(v))$
        \State Update embeddings:
        \State $h_v^k \leftarrow \sigma(\mathbf{W}^k \cdot \operatorname{CONCAT}(h_v^{k-1}, h_{\mathcal{N}(v)}^k))$
    \EndFor
\EndFor
\For{$v \in \mathcal{V}$}
    \State Generate final node embeddings: $z_v \leftarrow h_v^K$
\EndFor
\For{$uv \in \mathcal{E}$}
    \State Create edge embeddings: $z_{uv}^K \leftarrow \operatorname{CONCAT}(z_u^K, z_v^K)$
\EndFor
\State \textbf{return} $z_{uv}^K$, $\forall uv \in \mathcal{E}$ and $z_v$, $\forall v \in \mathcal{V}$
\end{algorithmic}
\end{algorithm}

E-GraphSAGE has been designed to address the identified gaps by making the necessary adjustments outlined in Algorithm 1. It is important to understand the differences between the base GraphSAGE \cite{Hamilton2017} and the evolved E-GraphSAGE, particularly in terms of algorithmic inputs, message propagation function, and output features.

The algorithm requires the inclusion of edge features $\left\{\mathbf{e}_{uv}, \forall {uv} \in \mathcal{E}\right\}$ for edge feature message propagation, absent in the original GraphSAGE. E-GraphSAGE uses only edge features to detect network intrusions since many intrusion detection datasets contain primarily edge information. While node features are typically initialized as constant vectors (e.g., $\mathbf{x}_{v} = \{1, \ldots, 1\}$ for each node), it's important to recognize that these constants may lack meaningful information. An alternative approach is to use meaningful representations such as vectors generated by Node2Vec.

To address the gaps in E-GraphSAGE, our study introduces two key enhancements: scattering transformations and Node2Vec initialization, which are absent in the standard E-GraphSAGE \cite{caville2022anomal}. The benefits and implications of these innovations are detailed in section \ref{steg}.

Neighborhood data aggregation shifts the focus from nodes to edge features. A proposed neighborhood aggregator function constructs aggregated edge features from sampled neighborhood edges, assimilating data from neighbors at the $k$-th layer.

\begin{equation}
\small
\label{eq:general_edge_aggregator}
\mathbf{h}_{\mathcal{N}(v)}^k \leftarrow \operatorname{AGG}_{k}\left(\{\mathbf{h}_u^{k-1}, \mathbf{e}_{uv}^{k-1}\}, \forall u \in \mathcal{N}(v)\right) .
\end{equation}

At layer $k$-$1$, edge features $\mathbf{e}_{uv}^{k-1}$ represent characteristics of edge $uv$ from $\mathcal{N}(v)$, the sampled neighborhood of node $v$. Effective message propagation through edge features involves combining these aggregated edge features with neighborhood node embeddings $\mathbf{h}_{u}^{k-1}$.

In line 13, the embedding for node $v$ at layer $k$ is derived from $k$-hop edge features, assimilating information from neighboring edges and $v$'s own features. This captures the topological patterns of the graph and the previous state of the node, possibly combined with the node representation for $v$ in the current node $\mathbf{h}_{v}^{k-1}$.

At depth $K$, the ultimate node representations are given by $\mathbf{z}_v = \mathbf{h}_v^K$, determined in line 17. For every edge, $uv$, the concluding edge embeddings $\mathbf{z}_{uv}^{K}$ are derived by concatenating the embeddings of nodes $u$ and $v$, as illustrated in Equation~(\ref{eq:final_edge_embedding}).

\begin{equation}
\small
\label{eq:final_edge_embedding}
\mathbf{z}_{uv}^{K} = {CONCAT}\left(\mathbf{z}_{u}^{K}, \mathbf{z}_{v}^{K}\right),~for~ uv \in \mathcal{E}.
\end{equation}
This signifies the culmination of the forward propagation phase within E-GraphSAGE.

\section{Methodology}\label{anoml-E}
 \subsection{Scattering Transform with E-GraphSAGE (STEG)}\label{steg}
 In this section, we will explore the intricacies of the Scattering Transform integrated with E-GraphSAGE, referred to as STEG. This combination aims to leverage the strengths of both techniques to provide a robust representation of data in the graph domain. While E-GraphSAGE effectively captures neighborhood information, the Scattering Transform supports multi-scale data analysis.
 
 \subsubsection{Scattering Transform}
 With its foundation in wavelet theory, the Scattering Transform provides a multi-scale breakdown method for feature vectors \cite{mallat2012st}. It provides a translation-invariant representation while capturing complex hierarchies within the data. Although the Scattering Transform is theoretically similar to deep convolutional networks, it is deterministic and works without training.

The mother and father wavelets are the essential components of the scattering transform. A well-defined function with zero mean, the mother wavelet, denoted by \(\psi\), provides a condensed representation in both the spatial and frequency domains. The breakdown of incoming signals or data is aided by the set it derives through dilation and translation. In contrast, the father wavelet, represented by \( \phi \), serves primarily as a smoothing function, pivotal for the low-pass filtering processes in the Scattering Transform.

\begin{figure}[!t]
    \centering
        \includegraphics[width=1\columnwidth]{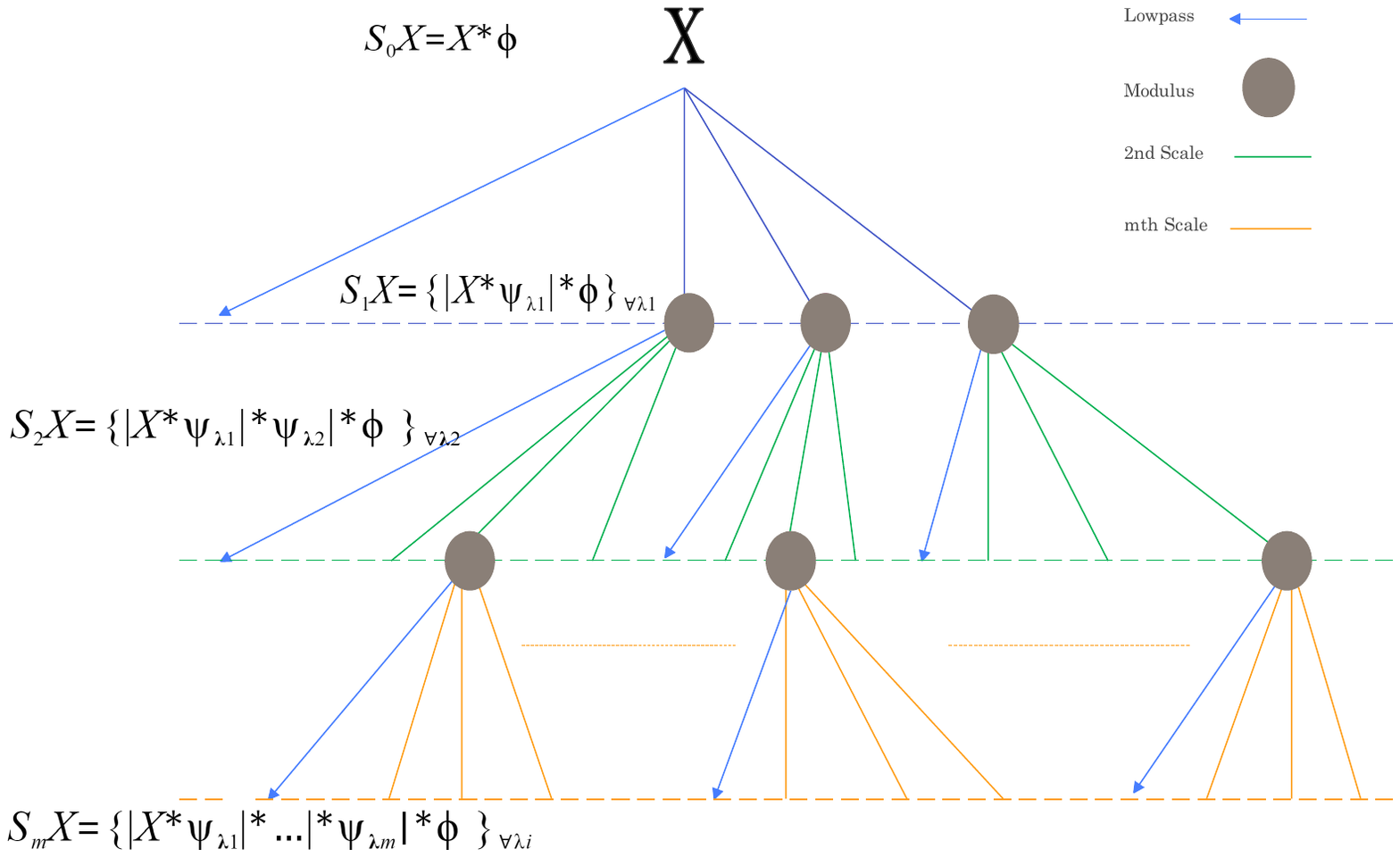}
    \caption{Multiscale Wavelet Decomposition Structure Analyzing Hierarchical Transformations and Convolution Operations on Signal X} 
    \label{fig:scattering_overview}
\end{figure}

Figure \ref{fig:scattering_overview} shows The Wavelet Scattering Transform represented by the input $X$, be it a signal, image, or feature vector,  as a multi-layered, translation-invariant representation. It proceeds through a sequence of modifications at various scales, denoted by different colored lines. At the outset, $X$ is convolved with the father wavelet \( \phi \), resulting in 
\begin{equation}
S_0X = X \ast \phi.
\end{equation} 
This operation primarily captures low-frequency components of the input. However, to capture and analyze higher frequencies, wavelet modulus transforms come into play.

\noindent The mother wavelet \( \psi \) provides a basis for analyzing $X$ at different scales or resolutions. The scale or level of the wavelet is controlled by the parameter \( \lambda \), which represents dilation. Specifically, \( \psi_{\lambda} \) denotes the mother wavelet dilated by a factor of \(\lambda \). As \(\lambda \) increases, the wavelet becomes wider, targeting lower frequencies.

To obtain the first-order scattering coefficient, the input \( X \) is first convolved with \( \psi_{\lambda} \) \cite{al2022feature}, and the modulus of the result is taken. This modulus output is then convolved with \( \phi \):
\begin{equation}
S_1 X=\left\{\left|X * \psi_{\lambda_1}\right| * \phi\right\}_{\forall \lambda_1}
\end{equation}

For wavelet scales where \(\lambda_1 \leq J\), it should be noted that \( J \) is a parameter that defines the maximum scale of the wavelet transformations. As we move to higher order coefficients, this process is iteratively applied. For instance, the second-order scattering coefficient employs wavelets at scales \( \lambda_1 \) and \( \lambda_2 \):
\begin{equation}
S_2X = \left| \left| X \ast \psi_{\lambda_1} \right| \ast \psi_{\lambda_2} \right| \ast \phi.
\end{equation}

As observed in the given figure, this structured hierarchy advances, with recurrent convolution and modulation operations at each scale. So, for the general  \( m \)-th order scattering coefficient, it can be represented as:

\begin{equation}
{S_m X(t)=\left\{\| X * \psi_{\lambda_1 }|* \ldots| * \psi_{\lambda_m }\mid  \ast \phi\right\}_{\forall  i} \quad \mathrm{i}=1,2, \ldots, \mathrm{m} .} 
\end{equation}

This mechanism ensures the capture of multi-scale features across a spectrum of frequencies. The choice of maximal order \( k \) is pivotal, influenced by the desired data granularity and computational feasibility. For many applications, a truncated hierarchy delivers the desired output without excessive computations.
\vspace{0.1cm}
\subsubsection{STEG}
\vspace{0.1cm}

\begin{figure*}[!th]
    \raggedleft 
        \includegraphics[width=\textwidth]{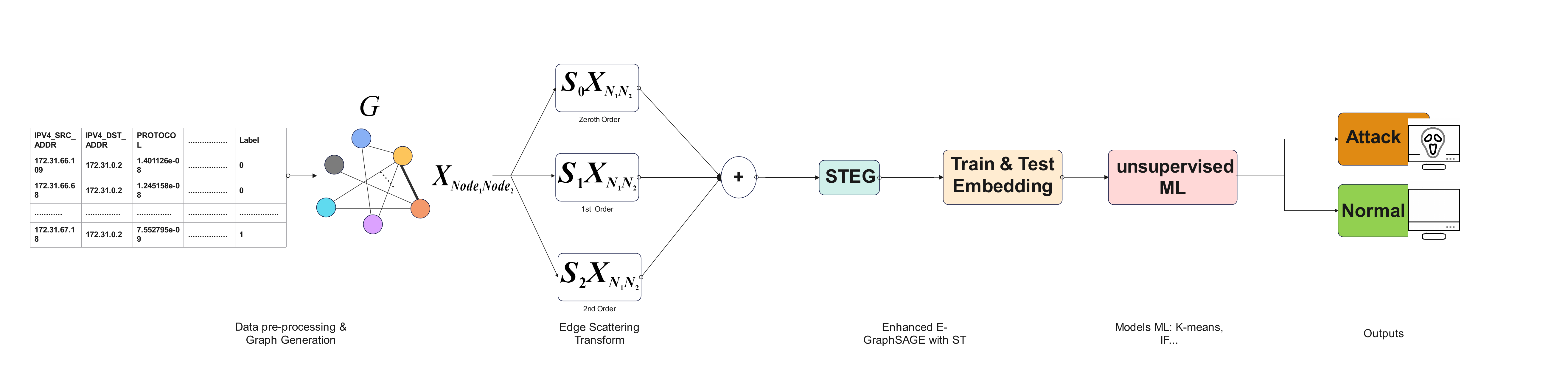}
    \caption{Multi-Order Scattering Transform Enhanced Graph-based Anomaly Detection: The procedure begins by converting raw IP-based network data into a graph representation. These enhanced embeddings are rigorously trained and tested using the Scattering Transform coupled with the E-GraphSAGE (STEG) method. Following training, the revised embeddings are directed into unsupervised machine learning approaches to classify network activity as ’Attack’ or ’Normal’.}
    \label{fig:proposed_method}
\end{figure*} 
\begin{figure}[!t]
    \raggedleft 
        \includegraphics[width=0.90\columnwidth]{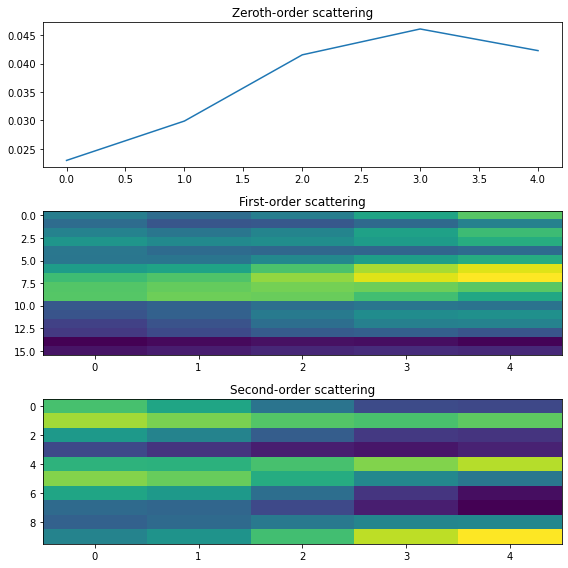}
    \caption{Visual Aptitude: Using Scattering Transforms to Show Multi-Layered Insights}
    \label{fig:frequency}
\end{figure}

In modern network analysis, nodes represent IP addresses that are the epicentres of computational interactions, while edges determine the nature and dynamics of those interactions. Each edge is defined by a sophisticated feature vector that encapsulates a wide range of network behaviours and properties.

To illustrate, individual notes in a piece of music can reveal the complexity of a melody or the contribution of a single instrument. However, the overall rhythm and tempo give an insight into the overall mood and genre of the song. Understanding these two scales is essential to fully appreciate the depth of a composition. Similarly, the scattering transform provides multi-resolution capability for network data, allowing a more nuanced understanding of both transient and long-term dynamics and overarching communication patterns. When applied to edge feature vectors, it facilitates robust multi-resolution exploration.

Due to its hierarchical data decomposition and ability to create invariant representations, the scattering transform emphasises the most discriminative features that are critical for anomaly detection. As a result, the scattering transform acts as a refined lens on the vast web of network connections, improving the clarity and precision with which anomalies are detected.

The first step of the process shown in figure \ref{fig:proposed_method} is to create a graph based on network flow, with nodes representing individual IP addresses. Edge features in the graph capture the interaction or flow between these addresses. To extract the rich information from these edges, the scattering transform is used. The zeroth order provides a global perspective, capturing the overarching properties of the edge features. The first order identifies primary variations in the data, such as specific anomalies or patterns in communication. The second order dives deeper, focusing on nuances within these primary patterns, revealing the full complexity of interactions. Once these multi-layered vectors are derived, they're concatenated with the original features, ensuring a thorough representation of edge information. This paves the way for robust data analysis. After generating embeddings from this enriched dataset, unsupervised machine learning models are used to classify network flows as either benign or indicative of potential attacks.

Focusing on the scattering transform stage, Figure \ref{fig:frequency} illustrates how the scattering transform captures minute details and complex patterns in an arbitrarily selected feature vector from the dataset NF-CSE-CIC-IDS2018-v2 \cite{icissp18}.
Moving to the first-order scattering heatmap, the intricate variations become evident. Each stripe and color variation in this heatmap corresponds to different flow patterns between machines; The more prominent, brighter patches may indicate regions of increased activity or potential inconsistencies in the flows. With this level of granularity, the chances of missing a subtle anomaly are greatly reduced. Finally, The second-order scattering, the bottom heatmap, further deepens our understanding. By examining the frequency of interactions, it showcases the complex relationships and dependencies in the flow data. The presence of noticeably brighter bands may indicate instances where multiple related events or factors culminate in an unusual flow pattern.

\subsection{Node2Vec initialization with E-GraphSAGE}
\begin{figure}[!t]
    \raggedleft 
        \includegraphics[width=1\columnwidth]{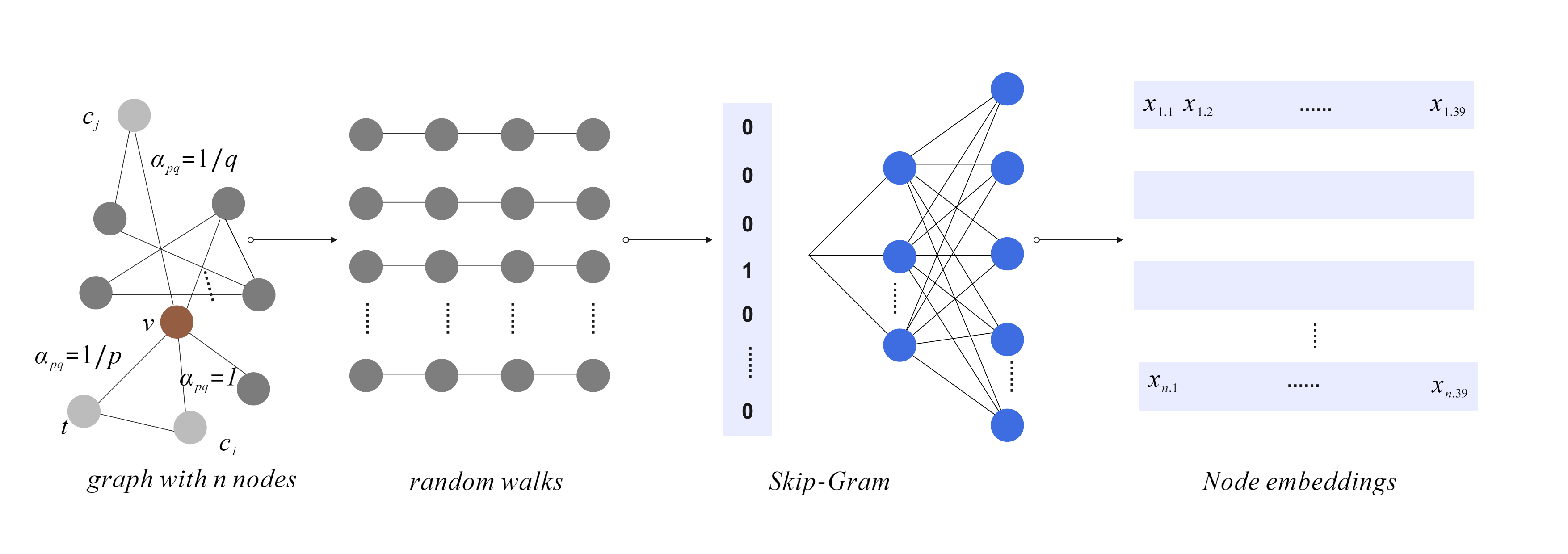}
    \caption{From Graph Topology to Vector Embeddings: The Node2Vec Transformation Process}
    \label{fig:N2V}
\end{figure}

The E-GraphSAGE approach \cite{lo2021graphsage,caville2022anomal} focuses exclusively on edges, frequently ignoring nodes in graphs by assigning them a constant value such as $\{1, \ldots, 1\}$. Our research tries to correct this by assigning additional logical values to these nodes. Figure  \ref{fig:N2V} depicts a transformation process in which a network with interconnected nodes is mapped into a binary matrix. The following matrix representation reveals whether or not connections exist. This matrix is then used to create feature vectors for individual nodes, which are denoted by  $\mathbf{x}_{1,1} ,\mathbf{x}_{1,2},... \mathbf{x}_{n,39}$ and so on.The number 39 ind Each node is endowed with a collection of properties that represent its distinct position and relationships within the graph.

Node2Vec is used as initialization and input for E-GraphSAGE. Node2Vec is a sophisticated technique for constructing node embeddings within a network. It integrates both structural and contextual aspects of the network to rapidly explore the large number of node neighbours. The procedure starts with random walks that start at each node and weave around the network to create sequences of nodes that reflect their immediate environment. These sequences are then used to train a skip-gram model \cite{qu2018node2defect}. This model in the Node2Vec context is primed to anticipate the nearby context of a node within the sequence. It borrows the essence from the famous Word2Vec methodology in natural language processing. As a result, each node has a rich, dense vector representation.

on the left side of Figure \ref{fig:N2V}, a fundamental aspect of Node2Vec is the random walk characterized by two parameters: $p$ and $q$ \cite{peng2019predicting}. Assume the random walk is currently at node v and the previous step was at node $t$. To determine the next position, it's imperative to evaluate the transition probabilities $\pi_{v c}$ on edges $(v, c)$ extending from $v$. The unstandardized transition probability is defined as $\pi_{v c}=\alpha_{p q}(t, c) \cdot w_{v c}$. Explicitly, $\alpha_{p q}$ is defined specifically as follows.

$$
\alpha_{p q}= \begin{cases}\frac{1}{p} & d_{t c}=0 \\ 1 & d_{t c}=1 \\ \frac{1}{q} & d_{t c}=2\end{cases}
$$
where $d_{t c}$ represents the shortest distance between nodes $t$ and $c$. 

For our anomaly detection dataset, we chose to set both $p$ and $q$ to 1. The rationale behind this choice is to maintain a balance between exploration and exploitation during the random walks. By equating $p$ and $q$ to 1, the random walk is neither biased towards immediately connected neighbours (exploitation) nor towards exploring further out in the graph (exploration). This balance ensures that the embeddings capture a more generalized representation of the nodes, which is crucial for anomaly detection where deviations from general patterns are of primary interest.

\subsection{Data Preparation and Network Graph Building}
First, we transform our dataset from the native NetFlow format \cite{rfc3954} into a graph-based representation to prepare for anomaly detection training. NetFlow collects IP flows and documents them for analysis \cite{caville2022anomal}. IP flows contain important network information such as IP addresses, packet counts, and statistics.

\begin{figure}[!th]
        \centering
        \includegraphics[width=0.98\columnwidth]{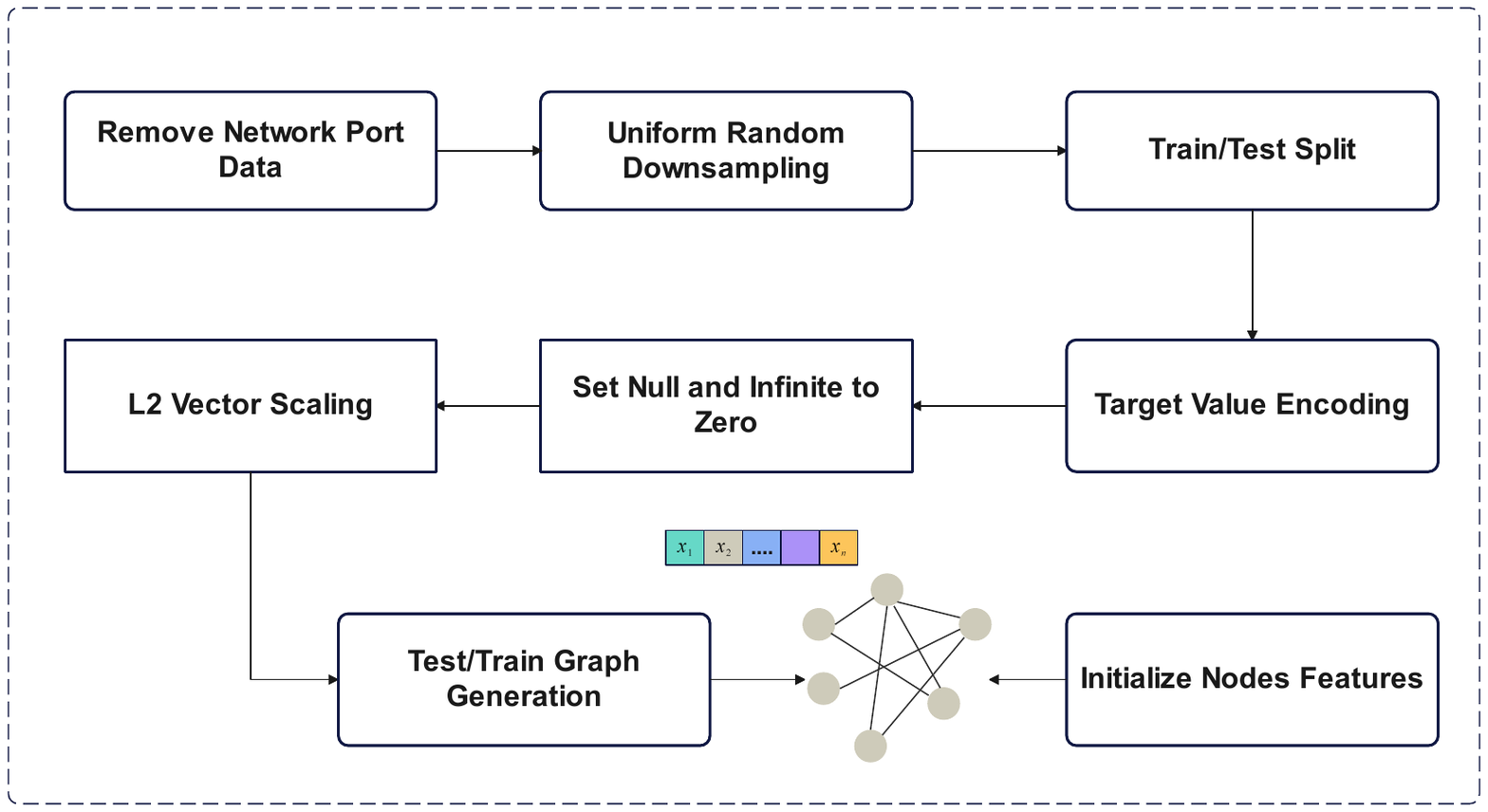}
    \caption{Data Preprocessing and Graph Generation Framework for Machine Learning.}
    \label{fig:preprocess}
\end{figure}

However, not all data elements are relevant to anomaly detection. Port numbers, as seen in Figure \ref{fig:preprocess}, although captured by NetFlow, are excluded because they don't affect anomaly detection. This refinement streamlines our data set and improves detection accuracy.
            
We represent IP addresses as nodes and traffic flow as edges, constructing a concise and meaningful graph for anomaly identification. To manage the complexity, we downsample the dataset to 10\% of its original size, following the approach of Singla et al. \cite{singla2020preparing}. This downsampling is critical for data management.
After downsampling, we divide the dataset into training and test sets and apply target encoding to categorical features. This unsupervised encoding ensures standardized data treatment. All empty or infinite values are replaced with zeros, which represent inactivity in the network traffic and aid in anomaly detection\cite{caville2022anomal}. This strategy is very useful in the context of anomaly detection. Zeros effectively represent a state of inactivity or absence in network traffic, providing a clear baseline from which abnormalities can be recognized. Following that, both the training and testing sets are $L_2$ normalized. The normalization parameters are obtained entirely from the training set, in keeping with our methodology's unsupervised approach. This assures that the normalization procedure is free of potential biases caused by the inclusion of test data.

 In the final stage of data processing, we convert the training and testing sets into bidirectional graph formats, enhancing the quality of our graph representations. While our methodology aligns with the E-GraphSAGE framework, we introduce unique approaches to initializing node attributes. Initially, we adopt the original E-GraphSAGE method, aligning node features' dimensions with edge features. Additionally, we incorporate the scattering transform technique to enrich our graph structure analysis (\ref{steg}). 
 This integration adds complexity and depth to our framework. Alternatively, we depart from the traditional method by initializing node features with Node2Vec embeddings. This strategy aims to capture more authentic and meaningful node information, resulting in a more realistic network representation, particularly beneficial for detailed topology analysis and inter-node relationships.

\subsection{Anomaly Detection Training Approach} 

In our research on network anomaly detection, training was systematically conducted using the generated training graph for each dataset and experimental setup. The machine learning models employed an encoder configured to E-GraphSAGE, commencing prior to the training phase. This encoder utilizes a mean aggregation function and is based on hyperparameters similar to those described \cite{lo2021graphsage,caville2022anomal}. Notably, our methodology diverges in its implementation of the STEG approach, where scattering is integrated prior to the aggregation function. Details of the hyperparameter selection are outlined in Table \ref{steg_parameters_left}.

In our methodology, a key focus was the empirical optimization of parameters for the Scattering1D transform, integral to our analysis. Through extensive experimentation, we determined the optimal settings to be \( J \) = 4 and \( Q \) = 16. The \( J \) parameter, defining the number of scales, and Q, indicating the number of wavelets per octave, were crucial in achieving the desired resolution for our analysis. This specific configuration proved effective in capturing the complexities and subtle nuances of network traffic and its associated anomalies. The \( J \) = 4 setting enabled us to conduct thorough investigations across a broad range of frequency scales, essential for identifying diverse anomalies. Concurrently, \( Q \) = 16 facilitated a detailed exploration within each octave, a critical factor for detecting fine-grained variations indicative of network anomalies.

Implementing this methodology produced 222 coefficients for each edge feature, resulting in a comprehensive and multidimensional representation of network traffic. This level of analysis improves our ability to detect and comprehend the wide range of anomalies present in the network. The encoder in the STEG approach employs a hidden layer of 256 units, which corresponds to the final output layer size. However, when generating edge embeddings, the size doubles to 512 units. The ReLU function is used for activation, and the model does not include a dropout rate.

We set the Node2Vec dimensionality to 64 for the second approach, which uses Node2Vec for node feature initialization while keeping the same configuration for the hidden layer as in the Anomal-E model (128 units), as shown in \ref{n2v_parameters_right}

In our approaches for network anomaly detection, we adopted a similar approach to the global graph summary generation as delineated in  Anomal-E \cite{caville2022anomal}, However, a notable deviation in our implementation is the exclusion of the Deep Graph Infomax (DGI) model's training loop. This decision was guided by our empirical findings that the training loop did not substantially improve the performance of the model in the context of our specific dataset and objectives. By omitting this training loop, we streamlined our model, focusing on leveraging the direct capabilities of the encoder to efficiently generate and utilize edge embeddings.
\begin{table}[!ht]
\small
\centering

\begin{minipage}{0.48\linewidth} 
\centering
\renewcommand{\arraystretch}{1.3}
\caption{Hyperparameter values used in STEG.}
\label{steg_parameters_left}
\scalebox{0.8}{
\begin{tabular}{*2l}
    \toprule
    Hyperparameter & Values \\
    \toprule
    No. Layers & 1 \\
    No. Scale \( J \) & 4 \\
    No. Wavelets per Octave \( Q \) & 16 \\
    No. Hidden & $256$ \\
    Activation Func. & ReLU \\
\end{tabular}
}
\end{minipage}
\hfill
\begin{minipage}{0.48\linewidth} 
\centering
\renewcommand{\arraystretch}{1.3}
\caption{Hyperparameter values used with Node2Vec initialization.}
\label{n2v_parameters_right}
\scalebox{0.8}{
\begin{tabular}{*2l}
    \toprule
    Hyperparameter & Values \\
    \toprule
    No. Layers & 1 \\
     Dimensions  & 64\\
    \( p , q \)  & 1\\
    No. Hidden & $128$\\
    Activation Func. & ReLU \\
\end{tabular}
}
\end{minipage}

\end{table}

\subsection{Implementing and Refining Anomaly Detection Algorithms}\label{anom-detect}

After training our models, namely \textit{STEG and Node2Vec initialization with E-GraphSAGE}, we generated edge embeddings from both training and test graphs for anomaly detection. Using PCA, IF, CBLOF, HBOS, and KMeans algorithms known for their effectiveness in unsupervised learning, we fed them edge embeddings from the training graph. Our experiments included contaminated and non-contaminated training samples to evaluate the performance of the algorithms under different conditions. Following the procedural framework of Anomal-E \cite{caville2022anomal}, we performed grid searches to fine-tune parameters for each algorithm, including contamination levels and specific settings such as the number of principal components for PCA, the number of estimators for IF, and the number of bins for HBOS. For KMeans in particular, we paid particular attention to adjusting the number of clusters to improve the effectiveness of data classification. These grid search parameters are detailed in the table \ref{grid_search}, reflecting our commitment to methodological precision.

\begin{table}[!ht]\small
\renewcommand{\arraystretch}{1.3}
\caption{Grid Search Hyperparameter Range for Anomaly Detection Algorithms.}
\label{grid_search}
\centering
\scalebox{0.8}{ 
\begin{tabular}{*3l }
    \toprule
     
    Algorithm & Hyperparameter & Values  \\
    \toprule
    K-MEANS & No. clusters & $[2, 3, 5, 7, 9, 10]$\\
    PCA & No. components & $[5, 10, 15, 20, 25, 30]$\\
    IF & No. estimators & $[20, 50, 100, 150]$\\
    CBLOF & No. clusters & $[2, 3, 5, 7, 9, 10]$\\
    HBOS & No. bins & $[5, 10, 15, 20, 25, 30]$\\
    \toprule
    All & contamination & $[0.001, 0.01, 0.04, 0.05, 0.1, 0.2]$\\
\end{tabular}
}
\end{table}

In our anomaly detection framework, KMeans and CBLOF played an important role by using clustering techniques to identify anomalies. KMeans clustering \cite{ahmed2020k} effectively grouped edge embeddings into clusters, distinguishing outliers by identifying data points significantly away from cluster centroids, thus exploiting spatial relationships within the data.

Complementing KMeans, the Cluster-Based Local Outlier Factor (CBLOF) algorithm \cite{he2003discovering} computed outlier scores for each data point, taking into account cluster size and distance from the nearest cluster center. This nuanced approach allowed CBLOF to provide detailed assessments of anomalies, capturing subtle variations in the data.

On the other hand, the PCA-based algorithm \cite{shyu2003novel} transformed high-dimensional edge embeddings into a lower-dimensional space, improving anomaly identification by highlighting deviations from normal patterns. In addition, the HBOS algorithm \cite{goldstein2012histogram} introduced a statistical method that constructs independent histograms for each feature to quickly detect outliers and efficiently process large datasets.

In addition, the Isolation Forest (IF) algorithm \cite{chabchoub2022depth} marks a significant departure from traditional techniques by isolating anomalies using random decision trees. This approach targets anomalies that are distinct and fewer in number, improving both the efficiency and effectiveness of anomaly detection. The integration of IF adds a distinctive element to our comprehensive network anomaly detection strategy.

\section{Experimental results}\label{results}
To evaluate the effectiveness of our methods, we adopted a rigorous testing approach using labelled datasets in both contaminated and non-contaminated forms. This allowed a comprehensive comparison with several benchmark anomaly detection algorithms. Due to the large size of the original datasets, they were uniformly reduced to $10\%$ of their original size. The training process used $70\%$ of this reduced data, with the remaining $30\%$ used for testing and in-depth performance analysis.
    
\subsection{Datasets}
We used two revised NetFlow formatted NIDS datasets for our analysis: NF-UNSW-NB15-v2 and NF-CSE-CIC-IDS2018-v2. These datasets, standardised by Sarhan et al. \cite{sarhan2022towards} are improved versions of pre-existing NIDS datasets, providing a consistent NetFlow format for better comparability.  In the non-contaminated experiments, training excluded all attack samples to maintain a pure dataset. Conversely, for the contaminated experiments, both datasets included 4\% of attack samples in the training phase. This specific level of contamination was selected to match the natural attack rate observed in NF-UNSW-NB15-v2, thereby ensuring consistent contamination levels across both datasets for a balanced and equitable experimental approach. More details about each dataset can be found below:
\vspace{0.1cm}
\subsubsection{NF-UNSW-NB15-v2}
\vspace{0.1cm}
This dataset is an updated NetFlow variant of the UNSW-NB15 dataset \cite{7348942}, which includes 43 standardized features. It contains a total of 2,390,275 network flows, of which 2,295,222 (96.02\%) are benign and 95,053 (3.98\%) are classified as attack flows, covering nine different attack types within the attack samples.
\vspace{0.1cm}
\subsubsection{NF-CSE-CIC-IDS2018-v2}   
\vspace{0.1cm}
This dataset comes from the CSE-CIC-IDS2018\cite{icissp18} and contains 18,893,708 flows. Of these, 16,635,567 (88.05\%) are benign, while 2,258,141 (11.95\%) are attack flows, with six different types of attacks among the attack samples.
       
\subsection{Evaluation metrics}\ \label{eval}
Our anomaly detection methodologies were compared to state-of-the-art techniques using macro F1 scores, highlighting their distinct benefits. The choice of macro F1 score, resilient to class imbalance, emphasized the improved performance of our approach with embeddings as inputs. Additionally, Accuracy (Acc) and detection rate (DR) were employed, confirming the superiority of our methods in the context of Network Intrusion Detection Systems (NIDS) over current leading methods.

    \subsection{STEG results}
    \subsubsection{NF-CSE-CIC-IDS2015-v2 Results }
        Table \ref{steg15} demonstrates the impressive efficacy of the STEG method when applied to the NF-UNSW-NB15-v2 dataset, highlighting its exceptional performance in uncontaminated and contaminated settings alike. Particularly noteworthy is STEG's outstanding performance in uncontaminated contexts, achieving a remarkable macro F1-score of 92.31\%. Even when the dataset is contaminated by 4\%, STEG maintains its leadership with an F1-score of 92.48\%. These findings surpass the performance of the Anomal-E approach, which attains F1 scores of 91.85\% in uncontaminated situations and 92.35\% in contaminated scenarios as best score.

\begin{table}[!ht]\small
            \renewcommand{\arraystretch}{1.3}
            \caption{NF-UNSW-NB15-v2 results using STEG }
            \label{steg15}
            \centering
            \scalebox{0.8}{
            \begin{tabular}{c | *3c | *3c}
                \toprule
                \multicolumn{1}{c}{} &  \multicolumn{3}{c}{0\% contamination} & \multicolumn{3}{c}{4\% contamination} \\ 
                {} & Acc & Macro F1 & DR & Acc & Macro F1 & DR \\
                KMeans & $98.68\%$ & $92.31\%$ & $96.07\%$ & $98.65\%$ & $92.14\%$ & $95.92\%$ \\
                PCA & $98.39 \%$ & $90.47\%$ & $90.60\%$ & $98.67\%$ & $92.42\%$ & $99.04\%$ \\
                IF & $98.43\%$ & $90.05\%$ & $83.71\%$ & $98.68\%$ & $92.47\%$ & $99.05\%$ \\
                CBLOF & $98.45\%$ & $90.21\%$ & $84.33\%$ & $98.61\%$ & $91.83\%$ & $94.18\%$ \\
                HBOS & $98.60\%$ & $91.88\%$ & $95.58\%$ & $98.69\%$ & $92.48\%$ & $98.04\%$ \\
            \end{tabular}
            }
        \end{table}

\begin{figure}[!ht]
  \centering

  \subfloat[\footnotesize Raw data]{\includegraphics[width=0.45\linewidth]{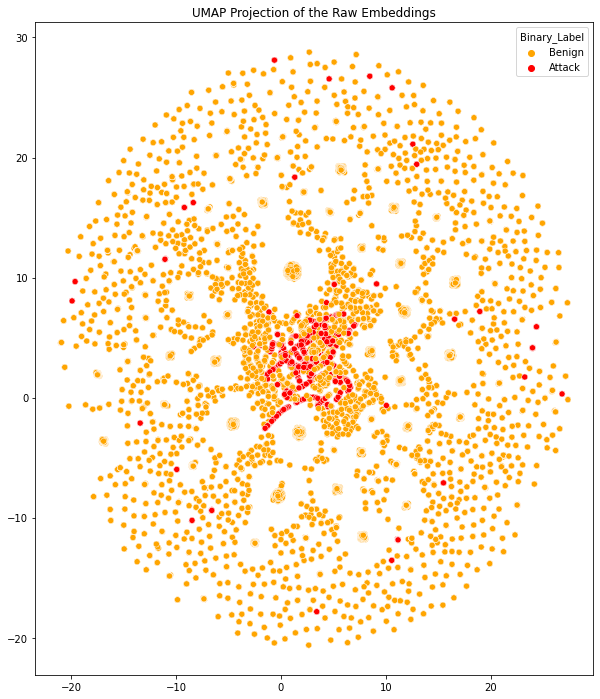}%
  \label{fig:rawdata}}
  \hfill
  \subfloat[\footnotesize STEG edge embedded data]{\includegraphics[width=0.45\linewidth]{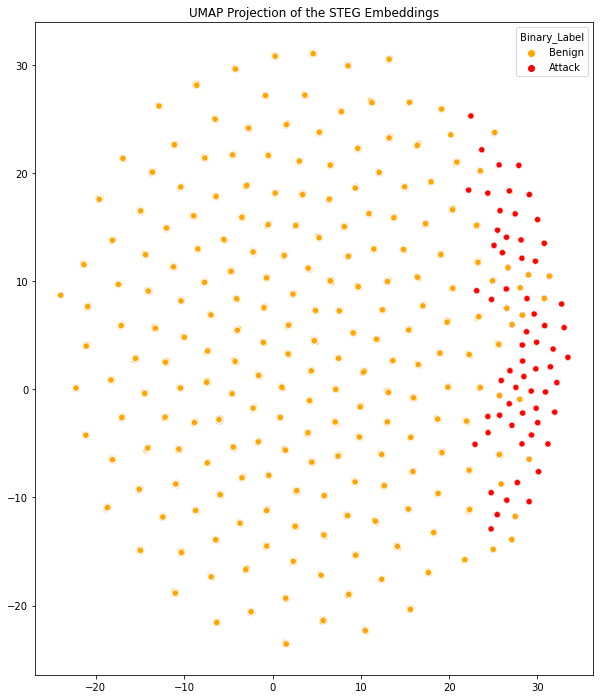}%
  \label{fig:comparison}}

  \caption{Comparison of raw and STEG edge embedded data visualization for NF-CSE-CIC-IDS2015-v2 dataset.}
  \label{fig:comparisonsteg}
  \captionsetup{font=scriptsize,labelformat=default} 
  \captionsetup[subfloat]{font=scriptsize,labelformat=simple} 
\end{figure}

In addition, the STEG method not only excels in quantitative metrics, but also significantly improves data preparation, as shown in Figure \ref{fig:comparisonsteg}. The visual representation of data embeddings transformed by STEG using Uniform Manifold Approximation and Projection (UMAP) shows a clear delineation between benign and attack data points, demonstrating a more defined separation compared to the raw data. This improved data structuring through STEG is critical to the enhanced performance of the anomaly detection algorithms, as evidenced by the consistent outperformance across all models compared to the Anomal-E method. As shown in the visual comparison, the graphical clarity of STEG's data embeddings demonstrates its effectiveness and creative method for preparing and separating complicated data for network anomaly detection.
       
\vspace{0.1cm}    
\subsubsection{NF-CSE-CIC-IDS2018-v2 Results}
\vspace{0.1cm}
Table \ref{steg18} further explores anomaly detection methodologies by showcasing the performance of the STEG approach on the NF-UNSW-NB18-v2 dataset, similar to its successful application in the NF-UNSW-NB15-v2 dataset. This comparative analysis highlights the effectiveness of the STEG method in two distinct data quality scenarios: one without contamination and the other with a 4\% contamination level.
Across the uncontaminated dataset, the STEG-CBLOF algorithm achieves an outstanding Macro F1 score of 95.71\% and exhibits an accuracy of 98.29\%, showcasing remarkable precision in anomaly identification. When subjected to a 4\% contamination rate, the resilience of STEG-CBLOF is evident through its sustained Macro F1 score of 94.60\% and minimal variation in detection rate, highlighting its adeptness in handling impure data.

        \begin{table}[!ht]\small
            \renewcommand{\arraystretch}{1.3}
            \caption{NF-UNSW-NB18-v2 results using STEG }
            \label{steg18}
            \centering
            \scalebox{0.8}{
            \begin{tabular}{c | *3c | *3c}
                \toprule
                \multicolumn{1}{c}{} &  \multicolumn{3}{c}{0\% contamination} & \multicolumn{3}{c}{4\% contamination} \\ 
                {} & Acc & Macro F1 & DR & Acc & Macro F1 & DR \\
                KMeans & $97.94\%$ & $94.72\%$ & $83.05\%$ & $90.74\%$ & $82.77\%$ & $94.89\%$ \\
                PCA & $98.15\%$ & $95.33\%$ & $85.35\%$ & $97.75\%$ & $94.58\%$ & $88.53\%$ \\
                IF & $97.78\%$ & $94.64\%$ & $88.53\%$ & $97.13\%$ & $93.24\%$ & $88.55\%$ \\
                CBLOF & $98.29\%$ & $95.71\%$ & $86.41\%$ & $97.89\%$ & $94.60\%$ & $82.64\%$ \\
                HBOS & $98.13\%$ & $95.29\%$ & $85.35\%$ & $97.52\%$ & $93.74\%$ & $82.65\%$ \\ 
            \end{tabular}
            }
        \end{table}

Moreover, the STEG-KMeans and STEG-PCA algorithms also demonstrate exceptional efficacy, with KMeans maintaining a notable increase in detection rate to 94.89\% under contamination—an insightful reflection of the algorithm's sensitivity to anomalous data. Similarly, the PCA variant retains its robust performance, achieving a commendable Macro F1 score of 95.33\% in a pristine environment and 94.58\% when subjected to contamination.
The Isolation Forest and HBOS algorithms, when integrated with the STEG framework, demonstrate a similar trend of marginal performance decline in the face of data contamination. The IF algorithm maintains a Macro F1 score well above 93\%, and HBOS shows a slight reduction to 93.74\%, yet both algorithms preserve high detection rates.

\subsection{Node2Vec initialisation  Results}
\subsubsection{NF-UNSW-NB15-v2 results }
\vspace{0.1cm}

Table \ref{cic_table_contamination} evaluates the efficacy of several anomaly detection algorithms initialized with Node2Vec embeddings on the NF-UNSW-NB15-v2 dataset, one with 0\% data contamination and the other with 4\% contamination. Results from Node2Vec initialization show consistent and strong performance across all algorithms. For example, under non-contaminated conditions, the KMeans algorithm achieves 98.67\% accuracy and a Macro F1 score of 92.22\%, while the PCA method demonstrates a slightly higher Macro F1 score of 93.92\% paired with an accuracy of 97.59\%. This trend of robust performance extends to the Isolation Forest (IF) and Histogram-based Outlier Score (HBOS) algorithms, which yield Macro F1 scores of 86.28\% and 92.30\%, respectively.
        
\begin{table}[!ht]\small
            \renewcommand{\arraystretch}{1.3}
            \caption{NF-UNSW-NB15-v2 results using Node2Vec }
            \label{cic_table_contamination}
            \centering
            \scalebox{0.8}{
            \begin{tabular}{c | *3c | *3c}
                \toprule
                \multicolumn{1}{c}{} &  \multicolumn{3}{c}{0\% contamination} & \multicolumn{3}{c}{4\% contamination} \\ 
                {} & Acc & Macro F1 & DR & Acc & Macro F1 & DR \\
                KMeans & $98.67\%$ & $92.22\%$ & $95.72\%$ & $98.65\%$ & $92.07\%$ & $94.71\%$ \\
                PCA & $97.59 \%$ & $93.92\%$ & $83.07\%$ & $97.77\%$ & $94.32\%$ & $82.84\%$ \\
                IF & $97,98\%$ & $86,28\%$ & $70.70\%$ & $98.72\%$ & $92,72\%$ & $1.00\%$ \\
                CBLOF & $98.71\%$ & $92.68\%$ & $1.00\%$ & $98.72\%$ & $92.74\%$ & $1.00\%$ \\
                HBOS & $98.65\%$ & $92.30\%$ & $98.21\%$ & $98.70\%$ & $92,64\%$ & $1.00\%$ \\
            \end{tabular}
            }
        \end{table}

In the presence of 4\% contamination, both Acc and Macro F1 scores show a slight overall decrease. However, the IF and CBLOF algorithms maintain Macro F1 scores above 92\%, demonstrating the robustness of Node2Vec-initialized embeddings with contaminated data.

Comparing these results to the Anomal-E suite of algorithms under similar conditions as reported in tables~\ref{state_art_no_contamination} and~\ref{state_art_contamination} in section~\ref{comp}, Node2Vec-initialized models exhibit competitive and sometimes superior performance. For example, the Node2Vec-initialized PCA algorithm outperforms Anomal-E-PCA's average Macro F1 score by 1.54\% under 0\% contamination and nearly 2\% under 4\% contamination. This highlights the potential benefits of combining Node2Vec embeddings with edge features to improve anomaly detection algorithms in network security datasets. In addition, Node2Vec-based algorithms maintain impressive detection rates (DR) even in contaminated environments. For example, when faced with a contamination level of 4\%, Node2Vec-initialized CBLOF and HBOS algorithms maintain a DR of 1.00\%, accurately pinpointing actual anomalies amidst cluttered data.

 \vspace{0.1cm}       
\subsubsection{NF-CSE-CIC-IDS2018-v2 Results}
\vspace{0.1cm}
        Similarly to NF-UNSW-NB15-v2 results, Table \ref{cic_table_contamination} presents the performance metrics of anomaly detection algorithms applied to the NF-UNSW-NB18-v2 dataset, utilizing the N2V-EGS method. Specifically, the CBLOF algorithm, when initialized with Node2Vec embeddings, demonstrates remarkable resilience. This is supported by a high Macro F1 score of 96.36\% under no contamination, gradually decreasing to 95.28\% with 4\% data contamination, indicating a robust response to data imperfections. The PCA technique also demonstrates significant consistency in performance, maintaining a high Macro F1 score of 96.16\% in a non contaminated state and 93.50\% when subjected to contamination, indicating a high level of algorithmic stability. The consistently high discrimination rates across these algorithms emphasize the role of the Node2Vec embedding in enhancing the algorithms' discriminatory capabilities.
        
        \begin{table}[!ht]\small
            \renewcommand{\arraystretch}{1.3}
            \caption{NF-UNSW-NB18-v2 results using Node2Vec }
            \label{cic_table_contamination}
            \centering
            \scalebox{0.8}{
            \begin{tabular}{c | *3c | *3c}
                \toprule
                \multicolumn{1}{c}{} &  \multicolumn{3}{c}{0\% contamination} & \multicolumn{3}{c}{4\% contamination} \\ 
                {} & Acc & Macro F1 & DR & Acc & Macro F1 & DR \\
                KMeans & $97.92\%$ & $94.68\%$ & $83.08\%$ & $97.53\%$ & $93.63\%$ & $80.54\%$ \\
                PCA & $98.46\%$ & $96.16\%$ & $87.90\%$ & $97.37\%$ & $93.50\%$ & $84.33\%$ \\
                IF & $97.91\%$ & $94.79\%$ & $85.58\%$ & $97.56\%$ & $93.95\%$ & $84.64\%$ \\
                CBLOF & $98.54\%$ & $96.36\%$ & $88.51\%$ & $98.06\%$ & $95.28\%$ & $88.52\%$ \\
                HBOS & $98.00\%$ & $95.14\%$ & $88.50\%$ & $97.50\%$ & $93.90\%$ & $85.94\%$ \\ 
            \end{tabular}
            }
        \end{table}
\begin{figure}[!ht]
  \centering

  \subfloat[\footnotesize Raw data ]{\includegraphics[width=0.45\linewidth]{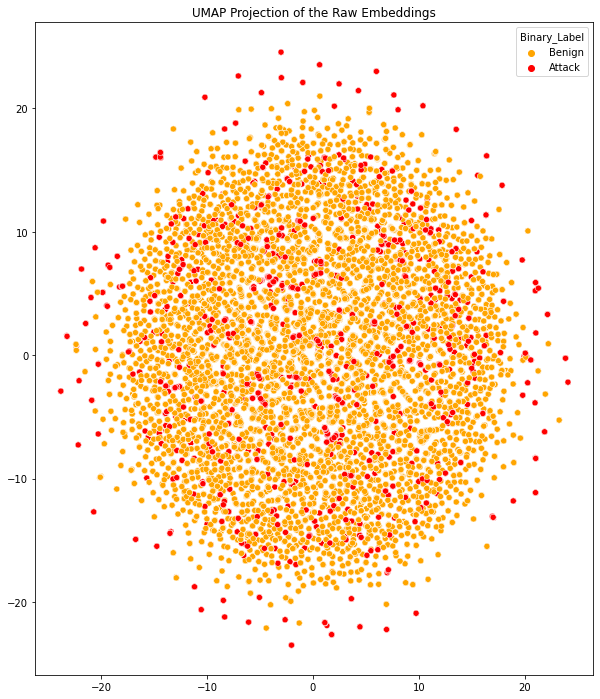}%
  \label{fig:rawdata}}
  \hfill
  \subfloat[\footnotesize N2V edge embedded data]{\includegraphics[width=0.45\linewidth]{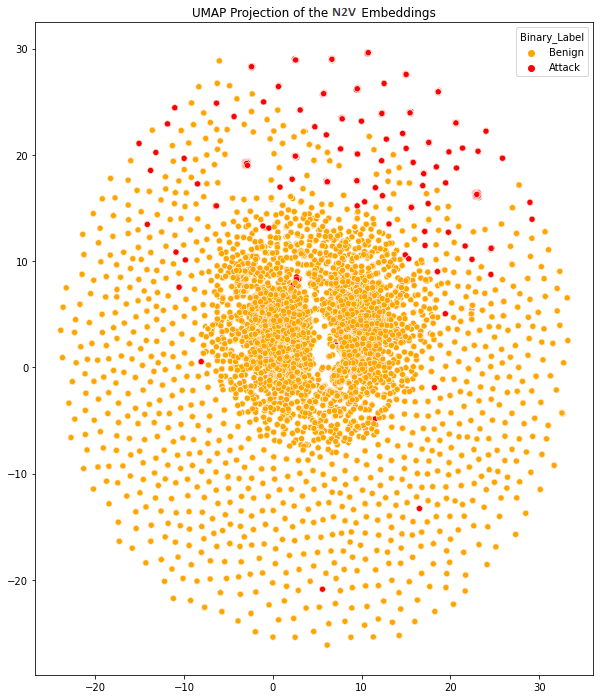}%
  \label{fig:comparison}}

  \caption{Comparison of raw and N2V-EGS edge-embedded data visualization for NF-CSE-CIC-IDS2018-v2 dataset}
  \label{fig:overallcomparison}
\end{figure}

Accompanying the tabulated data, the UMAP projections of the raw and Node2Vec edge-embedded data, as depicted in the corresponding Figure \ref{fig:comparison}, provide a clear visual representation of the classification abilities of the algorithms. The raw data visualization reveals a significant overlap between benign and malicious data points. In contrast, the Node2Vec edge-embedded data projection shows a more pronounced distinction between the two classes, enhancing the algorithms' capacity to accurately categorize and differentiate anomalies from normal instances.

\subsection{Comparative Evaluation of Anomaly Detection Enhancements}\label{comp}
In this section, we evaluate the performance of our proposed anomaly detection framework and compare it with several key baseline methods on the NF-UNSW-NB-15-v2 and NF-CSE-CIC-IDS2018-v2 datasets. Our framework introduces STEG, which combines the scattering transform with E-graphSAGE, and enhances the detection process by highlighting discriminative features through hierarchical data decomposition. By generating invariant representations, this innovative approach fine-tunes the focus on the complex web of network connections, significantly improving the accuracy of anomaly detection.

Furthermore, we have improved the E-GraphSAGE\cite{Lo2021EGraphSAGEAG} algorithm by incorporating node features, distinguishing our method from Anomal-E, which ignores the importance of nodes in graph analysis. This improvement to Node2Vec-EGS initialization is crucial for collecting a more comprehensive picture of network interactions. These upgraded STEG and Node2Vec-EGS embeddings have been used in a variety of anomaly detection techniques, as mentioned in Section ~\ref{anom-detect}. While we employed a number of techniques for baseline comparisons with GraphSAGE and DGI, our approach with KMeans is designed exclusively for a targeted comparison with Anomal-E. Tables \ref{state_art_no_contamination} and \ref{state_art_contamination}, along with Figure \ref{fig:nocont} present the results for non-contaminated and contaminated experiments, respectively, across both datasets.

In Table \ref{state_art_no_contamination}, our proposed methods, STEG and N2V-EGS, exhibited superior performance across all evaluated metrics on the NF-UNSW-NB15-v2 dataset. Specifically, the N2V-EGS-PCA variant achieved a Macro F1 score of 93.92\%, which is significantly higher by 10.33\% compared to Anomal-E and by approximately 45\% relative to DGI and GraphSAGE. This improvement illustrates the advantages of utilizing a comprehensive feature set that includes both edge and node features in the construction of embeddings for anomaly detection models. Additionally, the STEG method demonstrated a notable accuracy of 98.68\%, as illustrated in Figure \ref{fig:nocont}, surpassing baseline models. Furthermore, when tested on the NF-CSE-CIC-IDS2018-v2 dataset, our STEG and N2V-EGS methods continued to outperform, demonstrating significantly improved detection performance compared to the baseline methods in non-contaminated experiments. This underscores the resilience of our approaches, especially in scenarios involving clean data.

Demonstrating the generalizability of our methods is crucial, as evidenced by their consistent performance across various network attack datasets. Specifically, the STEG-HBOS and N2V-EGS-HBOS experiments serve as prime examples. Although the GraphSAGE-HBOS outperformed our HBOS-based methods marginally in the NF-CSE-CIC-IDS2018-v2 dataset, its average Macro F1 score across datasets dropped to 82.54\%. In contrast, our STEG and N2V-EGS approaches displayed improvement in this metric, achieving average Macro F1 scores of 93.59\% and 93.72\%, respectively. This pattern of reliability is evident across all variants of the STEG and N2V-EGS algorithms, affirming the effectiveness and adaptability of our proposed solutions in diverse anomaly detection scenarios.

In Table \ref{state_art_contamination}, the performance evaluations under conditions of 4\% contamination reveal that the proposed STEG and N2V-EGS methodologies consistently outperformed baseline algorithms across both datasets. Specifically, as depicted in Figure \ref{fig:nocont}  N2V-EGS-PCA achieved an exceptional Macro F1 score of 94.32\% on the NF-UNSW-NB15-v2 dataset and 93.50\% on the NF-CSE-CIC-IDS2018-v2 dataset, resulting in an impressive average of 93.91\% across datasets. Similarly, STEG-PCA demonstrated robust performance with an average Macro F1 score of 93.5\%. While demonstrating enhanced performance compared to the baseline, the Anomal-E suite of algorithms fell short of matching the high scores achieved by our methods. Specifically, Anomal-E-PCA achieved a respectable but somewhat lower average Macro F1 score of 92.38\%. In contrast, both DGI and GraphSAGE algorithms displayed noticeably inferior performance; with both DGI-PCA and GraphSAGE-PCA averaging Macro F1 scores below 48\%, clearly emphasizing the effectiveness of our proposed algorithms.

Further analysis indicates that our approaches demonstrate excellence not only in isolation but also in generalization across datasets. For instance, N2V-EGS-CBLOF and STEG-CBLOF achieved notably high average Macro F1 scores of 94.01\% and 93.22\% respectively. In contrast, GraphSAGE-CBLOF only attained an average of 72.07\%, indicating a lack of robustness in generalizing across datasets. A similar trend is observed in the HBOS variations, with N2V-EGS-HBOS and STEG-HBOS achieving averages of 93.27\% and 93.11\%, which are significantly higher than the 74.69\% average of GraphSAGE-HBOS. The results suggest that although certain traditional methods may function satisfactorily with specific datasets, they lack the ability to perform effectively in more diverse applications. In contrast, our STEG and N2V-EGS algorithms exhibit both strong performance and adaptability in scenarios involving contaminated data.

         \begin{table*}[!ht]
    \centering
    \includegraphics[width=0.92\textwidth]{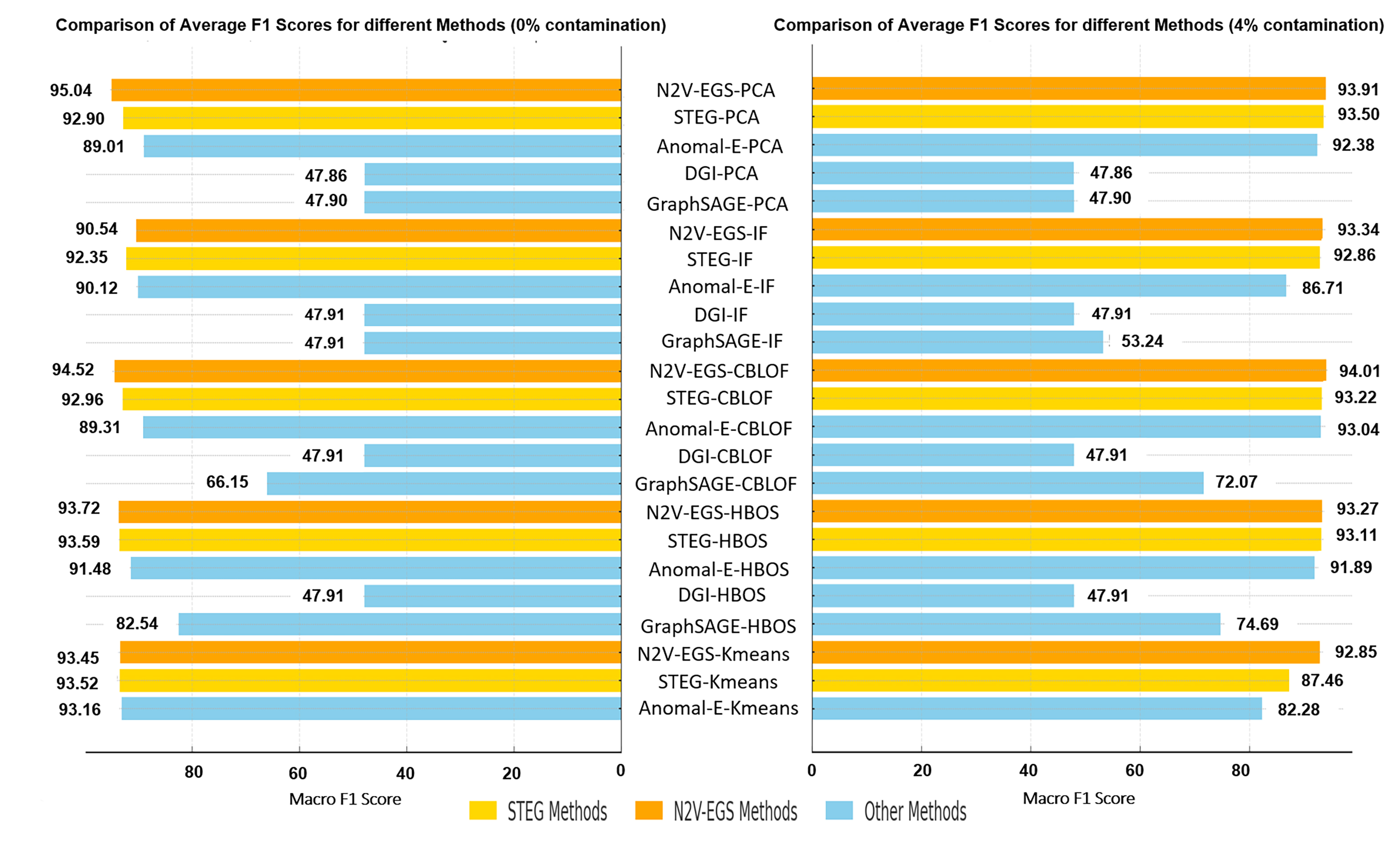}
    \captionof{figure}{Comparative Analysis of Average F1 Scores for Anomaly Detection Algorithms Under 0\% and 4\% Data Contamination}
    \label{fig:nocont}
\end{table*}
         \begin{table*}[!ht]
            \renewcommand{\arraystretch}{1.2}
            \caption{Performance evaluation for metrics on both datasets compared with the baseline algorithms (0\% contamination).}
            \label{state_art_no_contamination}
            \centering
            \scalebox{0.92}{
            \begin{tabular}{c | *3c | *3c | *1c}
                \toprule
                \multicolumn{1}{c}{} &  \multicolumn{3}{c}{NF-UNSW-NB15-v2} & \multicolumn{3}{c}{NF-CSE-CIC-IDS2018-v2} &
                \multicolumn{1}{c}{Average Across Datasets} \\ 
                {} & Acc & Macro F1 & DR & Acc & Macro F1 & DR & Macro F1 \\

                N2V-EGS-PCA (Ours) & $97.59\%$ &  $\textbf{93.92\%}$ & $\textbf{83.07\%}$ & $\textbf{98.46\%}$ & $\textbf{96.16\%}$ & $\textbf{87.90\%}$ &  \textbf{95.04\%}  \\
                
                STEG-PCA (Ours) & $\textbf{98.39\%}$ &  $\textbf{90.47\%}$ & $\textbf{90.60\%}$ & $\textbf{98.15\%}$ & $\textbf{95.33\%}$ & $\textbf{85.35\%}$ &  \textbf{92.90\%}  \\

                Anomal-E-PCA  & $97.64\%$ &  $83.59\%$ & $64.20\%$ & $97.82\%$ & $94.43\%$ & $82.67\%$ &  89.01\% \\

                DGI-PCA & $96.02\%$ & $48.89\%$ & $0.00\%$ & $88.03\%$ & $46.82\%$ & $0.00\%$ & 47.86\% \\
                GraphSAGE-PCA & $95.95\%$ & $48.97\%$ & $0.00\%$ & $88.05\%$ & $46.82\%$ & $0.00\%$  & 47.90\%   \\

\hline
                N2V-EGS-IF (Ours) & $\textbf{97.98\%}$ &  $\textbf{86.28\%}$ & $\textbf{70.70\%}$ & $\textbf{97.91\%}$ & $\textbf{94.79\%}$ & $\textbf{85.58\%}$ &  \textbf{90.54\%}  \\
                
                STEG-IF (Ours) & $\textbf{98.43\%}$ &  $\textbf{90.05\%}$ & $\textbf{83.71}\%$ & $97.78\%$ & $\textbf{94.64\%}$ & $\textbf{88.53\%} $ &  \textbf{92.35\%}  \\

                 Anomal-E-IF  & $97.91\%$ & $85.62\%$ & $68.76\%$ & $97.87\%$ & $94.56\%$ & $82.87\%$ &  90.12\% \\

                DGI-IF & $96.02\%$ & $48.99\%$ & $0.00\%$ & $88.03\%$ & $46.82\%$ & $0.00\%$ &  47.91\% \\
                GraphSAGE-IF & $96.02\%$ & $48.99\%$ & $0.00\%$ & $88.05\%$ & $46.82\%$ & $0.00\%$ &  47.91\% \\

\hline

                 N2V-EGS-CBLOF (Ours) & $\textbf{98.71\%}$ &  $\textbf{92.68\%}$ & $\textbf{1.00\%}$ & $\textbf{98.54\%}$ & $\textbf{96.36\%}$ & $88.51\%$ &  \textbf{94.52\%}  \\
                 
                 STEG-CBLOF (Ours) & $\textbf{98.45\%}$ &  $\textbf{90.21\%}$ & $\textbf{84.33}\%$ & $\textbf{98.29\%}$ & $\textbf{95.71\%}$ & $86.41\% $ &  \textbf{92.96\%}  \\
                 
                 Anomal-E-CBLOF  & $97.70\%$ & $84.17\%$ & $65.97\%$ & $97.83\%$ & $
                 94.44\%$ & $82.67\%$  & 89.31\% \\
                DGI-CBLOF & $96.02\%$ & $48.99\%$ & $0.00\%$ & $88.03\%$ & $46.82\%$ & $0.00\%$  & 47.91\% \\
                 
                GraphSAGE-CBLOF & $87.10\%$ & $49.53\%$ & $10.32\%$ & $90.74\%$ & $82.76\%$ & $\textbf{95.08\%}$  & 66.15\% \\
                
                \hline

                N2V-EGS-HBOS (Ours) & $\textbf{98.65\%}$ &  $\textbf{92.30\%}$ & $\textbf{98.21\%}$ & $98.00\%$ & $95.14\%$ & $88.50\%$ &  \textbf{93.72\%}  \\

                STEG-HBOS (Ours) & $\textbf{98.60\%}$ &  $\textbf{91.88\%}$ & $\textbf{95.58}\%$ & $98.13\%$ & $95.29\%$ & $85.35\% $ &  \textbf{93.59\%}  \\
                
                 Anomal-E-HBOS  & $98.18\%$ & $88.45\%$ & $80.36\%$  & $97.72\%$ & $94.51\%$ & $88.55\%$ & 91.48\%\\
                DGI-HBOS & $96.02\%$ & $48.99\%$ & $0.00\%$ & $88.03\%$ & $46.82\%$ & $0.00\%$ &  47.91\% \\
                GraphSAGE-HBOS & $96.91\%$ & $68.39\%$ & $24.22\%$ & $\textbf{98.67\%}$ & $\textbf{96.69\%}$ & $\textbf{88.94\%}$ &  82.54\%\\
                \hline
                N2V-EGS-Kmeans (Ours) & $\textbf{98.67\%}$ &  $\textbf{92.22\%}$ & $\textbf{95.72\%}$ & $\textbf{97.92\%}$ & $\textbf{94.68\%}$ & $\textbf{83.08\%}$ &  \textbf{93.45\%}  \\

                STEG-kmeans (Ours) & $\textbf{98.68\%}$ &  $\textbf{92.31\%}$ & $\textbf{96.07\%}$ & $\textbf{97.94\%}$ & $\textbf{94.72\%}$ & $\textbf{83.05\%} $ &  \textbf{93.52\%}  \\
                
                 Anomal-E-kmeans  & $98.62\%$ & $91.85\%$ & $94.14\%$  & $97.84\%$ & $94.47\%$ & $82.66\%$ & 93.16\%\\

            \end{tabular}
            }
            \label{state_art_no_contamination}
        \end{table*}

         \begin{table*}[!ht]
            \renewcommand{\arraystretch}{1.3}
            \caption{Performance evaluation for metrics on both datasets compared with the baseline algorithms (4\% contamination).}
            \label{state_art_contamination}
            \centering
            \scalebox{0.92}{
           \begin{tabular}{c | *3c | *3c | *1c}
                \toprule
                \multicolumn{1}{c}{} &  \multicolumn{3}{c}{NF-UNSW-NB15-v2} & \multicolumn{3}{c}{NF-CSE-CIC-IDS2018-v2} &
               \multicolumn{1}{c}{Average Across Datasets}\\ 
                {} & Acc & Macro F1 & DR & Acc & Macro F1 & DR  & Macro F1 \\

                N2V-EGS-PCA(Ours) & $\textbf{97.77\%}$ &  $\textbf{94.32\%}$ & $82.84\%$ & $\textbf{97.37\%}$ & $\textbf{93.50\%}$ & $\textbf{84.33\%}$ &  \textbf{93.91\%}  \\

                STEG-PCA (Ours) & $\textbf{98.67\%}$ &  $\textbf{92.42\%}$ & $\textbf{99.04\%}$ & $\textbf{97.75\%}$ & $\textbf{94.58\%}$ & $\textbf{88.53\%}$ &  \textbf{93.5\%}  \\
                
                Anomal-E-PCA  & $98.63\%$ & $92.18\%$ & $97.86\%$ & $97.11\%$ & $92.57\%$ & $79.16\%$ & 92.38\% \\

                DGI-PCA & $96.02\%$ & $48.89\%$ & $0.00\%$ & $88.03\%$ & $46.82\%$ & $0.00\%$ &  47.86\%  \\
                GraphSAGE-PCA & $95.95\%$ & $48.97\%$ & $0.00\%$ & $88.02\%$ & $46.82\%$ & $0.00\%$ & 47.90\%  \\

\hline
                N2V-EGS-IF(Ours) & $\textbf{98.72\%}$ &  $\textbf{92, 72\%}$ & $\textbf{1.00\%}$ & $\textbf{97.56\%}$ & $\textbf{93.95\%}$ & $84.64\%$ &  \textbf{93.34\%}  \\

                STEG-IF (Ours) & $\textbf{98.68\%}$ &  $\textbf{92.47\%}$ & $\textbf{99.05\%}$ & $\textbf{97.13\%}$ & $\textbf{93.24\%}$ & $88.55\%$ &  \textbf{92.86\%}  \\

                 Anomal-E-IF  & $98.66\%$ & $92.35\%$ & $98.77\%$ & $89.79\%$ & $81.11\%$ & $\textbf{91.84\%}$ & 86.71\% \\

                DGI-IF & $96.02\%$ & $48.99\%$ & $0.00\%$ & $88.03\%$ & $46.82\%$ & $0.00\%$ &  47.91\% \\
                GraphSAGE-IF & $60.55\%$ & $39.92\%$ & $24.57\%$ & $76.40\%$ & $66.56\%$ & $92.72\%$  &  53.24\% \\

\hline
                N2V-EGS-CBLOF(Ours) & $\textbf{98.72\%}$ &  $\textbf{92, 74\%}$ & $\textbf{1.00\%}$ & $\textbf{98.06\%}$ & $\textbf{95.28\%}$ & $\textbf{88.52\%}$ &  \textbf{94.01\%}  \\

                STEG-CBLOF (Ours) & $\textbf{98.61\%}$ &  $\textbf{91.83\%}$ & $\textbf{94.18\%}$ & $97.89\%$ & $94.60\%$ & $82.64\%$ &  \textbf{93.22\%}  \\

                 Anomal-E-CBLOF  & $98.57\%$ & $91.70\%$ & $95.72\%$ & $97.80\%$ & $94.38\%$ & $82.67\%$ &  93.04\% \\
                DGI-CBLOF & $96.02\%$ & $48.99\%$ & $0.00\%$ & $88.03\%$ & $46.82\%$ & $0.00\%$ & 47.91\% \\
                GraphSAGE-CBLOF & $87.10\%$ & $49.53\%$ & $10.32\%$ & $97.90\%$ & $94.61\%$ & $82.60\%$  &  72.07\%\\
                
\hline                

                N2V-EGS-HBOS(Ours) & $\textbf{98.70\%}$ &  $\textbf{92, 64\%}$ & $\textbf{1.00\%}$ & $97.50\%$ & $93.90\%$ & $\textbf{85.94\%}$ &  \textbf{93.27\%}  \\

                STEG-HBOS (Ours) & $\textbf{98.69\%}$ &  $\textbf{92.48\%}$ & $\textbf{98.04\%}$ & $97.52\%$ & $93.74\%$ & $82.65\%$ &  \textbf{93.11\%}  \\
                
                 Anomal-E-HBOS  &$98.62\%$ & $91.89\%$ & $94.92\%$   & $96.86\%$ & $91.89\%$ & $77.79\%$& 91.89\%  \\
                DGI-HBOS & $96.02\%$ & $48.99\%$ & $0.00\%$ & $88.03\%$ & $46.82\%$ & $0.00\%$  & 47.91\%  \\
                GraphSAGE-HBOS & $88.60\%$ & $54.77\%$ & $26.63\%$ & $\textbf{97.90\%}$ & $\textbf{94.61\%}$ & $82.60\%$ & 74.69\% \\

\hline                
                
               N2V-EGS-Kmeans (Ours) & $\textbf{98.65\%}$ &  $\textbf{92.07\%}$ & $\textbf{94.71\%}$ & $\textbf{97.53\%}$ & $\textbf{93.63\%}$ & $\textbf{80.54\%}$ &  \textbf{92.85\%}  \\

                STEG-kmeans (Ours) & $\textbf{98.65\%}$ &  $\textbf{92.14\%}$ & $\textbf{95.92\%}$ & $\textbf{90.74\%}$ & $\textbf{82.77\%}$ & $\textbf{94.89\%} $ &  \textbf{87.46\%}  \\
                
                 Anomal-E-kmeans  & $98.57\%$ & $91.64\%$ & $94.58\%$  & $85.45\%$ & $72.91\%$ & $72.69\%$ & 82.28\%\\

                \\
            \end{tabular}
            }
        \end{table*}

\section{Conclusion}\label{conclusion}
In summary, this paper presents groundbreaking methods in NIDS (NIDS) using Graph Neural Networks (GNNs), focusing on two advanced approaches: Scattering Transform with E-GraphSAGE (STEG) and Node2Vec initialization within the E-GraphSAGE framework \cite{Lo2021EGraphSAGEAG}. STEG, a novel combination of Scattering Transform and E-GraphSAGE, excels in multi-resolution analysis of edge feature vectors, improving the detection of subtle network anomalies, and represents a first in the use of self-supervised GNNs with Scattering Transform for network security. The Node2Vec approach, which differs from traditional techniques, improves anomaly detection by embedding real-world attributes into node representations. Our extensive experimental evaluations on NIDS benchmark datasets demonstrate the superiority of these methods over existing techniques and highlight their effectiveness against modern cyber threats. Future efforts will focus on refining these methods, exploring their adaptability to different network environments, and integrating them into broader cybersecurity strategies, thereby advancing GNN-based network intrusion detection and setting new precedents in technology-driven cybersecurity.

\end{document}